\documentclass{aa}

\usepackage{graphicx}
\usepackage{txfonts}
\usepackage{aalongtable}

\begin{document}
 


\title{Properties of RR Lyrae stars in the inner regions of the Large Magellanic Cloud. II.The extended sample.
\thanks{Based on observations collected with the 
Very Large Telescope, the New Technology Telescope
of the European Southern Observatory and GeminiS of Gemini Observatory
within the Observing Programs
64.N-0176(B),70.B-0547(A), 072.D-0106(B), GS-2004A-Q-27.
Table 9 is available in electronic form
at the CDS.}}

\author{J.\ Borissova\inst{1}
\and
D.\ Minniti\inst{2}
\and
M.\ Rejkuba\inst{3} 
\and
D. Alves\inst{4}
} 
\offprints{J.\ Borissova}

\institute{Departamento de F\'isica y Meteorolog\'ia, Facultad 
       de Ciencias, Universidad de Valpara\'{\i}so, 
       Ave. Gran Breta\~na 644, Playa Ancha, Casilla 53,
             Valpara\'iso, Chile, 
\email{jura.borissova@uv.cl}   
 \and     
      Department of Astronomy, P. Universidad Cat\'olica, 
      Av. Vicu\~na Mackenna 4860, Casilla 306, Santiago 22, Chile\\
      \email{dante@astro.puc.cl}
 \and
      European Southern Observatory, Karl-Schwarzschild-Str. 2, D-85748
 Garching b.  M\"{u}nchen, Germany\\
      \email{mrejkuba@eso.org}
 \and
   3549 Lynne Way, Sacramento, CA 95821-3722, USA\\
      \email{alvesdavid@ucdavis-alumni.com}}

\date{Received .. ... 2005; accepted .. ... 2005}

\authorrunning{Borissova et al.}
\titlerunning{Properties of LMC RR Lyrae stars}

\abstract
{All galaxies that have been adequately examined so far have shown an extended 
stellar halo.} 
{To search for such a halo in the LMC we have obtained low-resolution
spectra for 100 LMC RR Lyrae stars, of which 87 are in the field and 13 in 
the clusters NGC1835 and NGC2019.} 
{We measured radial velocities for 87 LMC RR Lyrae stars, 
and metallicities for 78 RR Lyrae stars, nearly tripling the previous 
sample. These targets are located in 10 fields covering a wide 
range of distances, out to 2.5 degrees from the center of the LMC.
} 
{Our main result is that the mean velocity 
dispersion for the LMC RR Lyrae stars is 
$\sigma_{RV} = 50 \pm 2$ km/s. This quantity does not appear 
to vary with distance from the LMC center. 
The metallicity shows a Gaussian distribution, with mean
[Fe/H]= -1.53 $\pm 0.02$ dex, and dispersion $\sigma_{[Fe/H]} = 
0.20\pm 0.02$ dex in the Harris  metallicity scale, 
confirming that they represent a very 
homogeneous metal-poor population. There is no dependence between the 
kinematics and metallicity of the field RR Lyrae star population.}
{Using good quality low-resolution spectra from FORS1, FORS2 and GEMINI-GMOS
we have found that field RR Lyrae stars in the LMC show a large
velocity dispersion and that this indicate the presence of 
old and metal-poor stellar halo. All the evidence so far for the halo,
however, is from the spectroscopy of the inner LMC regions, 
similar to the inner flattened halo in our Galaxy. 
Further study is necessary to confirm this important result.} 

 
\maketitle

\section{Introduction}

The LMC is an irregular, disk dominated, bulge-less, nearly face-on galaxy. 
It is of fundamental 
importance for studies of stellar populations and of the interstellar medium.
It is being used to study the presence of dark objects in the Galactic 
Halo through microlensing (e.g., Alcock et al. 2000a,b), 
it is an important 
probe of the Galaxy's formation history, and it plays a 
key role in determinations of the cosmological distance scale 
(e.g., Freedman et al. 2001).

All galaxies that have been adequately examined so far have shown an extended 
star halo. Furthermore the Lambda CDM cosmology predicts a dark matter halo for 
galaxies like the LMC. Unfortunately, outlining a stellar halo in the LMC 
(and its metal and age distribution), a key 
observation for galaxy formation scenarios, is made difficult by the 
large extent of this galaxy on the sky and inclination, causing the halo to
be projected against the LMC disk.

Recently, using low resolution spectroscopy with the
FORS1 spectrograph at the 
Very Large Telescope (VLT), Minniti et al. (2003) 
and Borissova et al. (2004; hereafter Paper~I) 
on the basis of  the radial velocities of 
43 RR Lyrae stars obtained a true velocity dispersion of 
53 km/s, which is higher than the  velocity dispersion of any other LMC 
population previously measured. This high  velocity dispersion
indicates that a kinematically hot metal-poor old halo exists in the LMC.  

Cole at al. (2005) reported metallicities and radial velocities derived 
from spectra at the near-infrared calcium triplet for 373 red giants in a 
200 arcmin$^2$ area at the optical center of the LMC bar. They found that 
the velocity dispersion of the whole sample is $24.7\pm0.4$ km~s$^{-1}$. 
When cut by metallicity, the most metal-poor 5$\%$ of stars 
($\mathrm{[Fe/H]} < -1.15$) show $40.8\pm1.7$ km~s$^{-1}$, 
more than twice the value for the most metal-rich 
suggesting that an old, thicker disk or halo population is present, in agreement
with our results from RR Lyrae stars.

The counter-argument against the evidence of a spherical halo in the outer
regions of LMC was presented by Alves (2004a,b), who re-examined the content 
of the 2MASS database at its periphery. 
Based on the dereddened colors of the LMC's red giant branch stars, 
he found an outward radial gradient of decreasing metallicity. The isodensity 
contours of the RGB stars rule out the existence of a spherical halo 
at the periphery, and reveal instead an inclined disk. 
Similar evidence, again from photometric data, was presented by 
Gallart et al.\ (2004) from a high-quality color-magnitude diagram 
(CMD) for a $36 \times 36$ arcmin field located at 7 kpc 
from the LMC center. The surface brightness profile of the LMC remains 
exponential to this large galactocentric radius and shows no evidence 
of disk truncation. Combining the information on surface brightness 
and stellar population, they concluded that the LMC disk extends 
(and dominates over a possible stellar halo) out to a distance of 
at least 7 kpc. 
Clementini et al. (2003) and Gratton et al. (2004) presented 
new photometry 
and spectroscopy for more than a hundred RR Lyrae stars in two 
fields located close to the bar of the LMC. 
Their average magnitude, the local reddening, individual metallicities, 
the luminosity-metallicity relation, 
and the distance to the LMC were derived. 
They calculated the dispersion of the mean radial velocity of the 48 stars with multiple 
observations of 40 km/s.  

Subramaniam (2006) investigated the distribution of 
the RR Lyrae star population in the inner LMC and found a halo-like location 
and disc-like density distribution. He also found that RR Lyrae stars in the 
inner LMC are elongated more along the line of sight than 
along the major axis of the disk. Thus there are many indications
that the LMC has a halo, which is similar to the inner halo of our Galaxy,
but the results are puzzling.
The outer halo of the LMC, which is characterized by a spatially 
extended old population with large velocity dispersion, has still not been
detected.  

In this paper we present new spectroscopic data extending
our sample to 137 field RR Lyrae stars and 13 RR Lyrae stars 
in the LMC globular clusters NGC1835 and NGC2019.
The next section describes the data and the third section discusses the
radial velocities of the extended sample. The fourth section includes the 
metallicities and the last 
section is a summary of the results.


\section{Observations and reductions}
 
In Paper~I we presented radial velocities, metallicities and $K$-band
magnitudes of 74 RR Lyrae stars in the inner regions of the LMC. 
The optical low resolution spectra were obtained with the FORS1 at the ESO 
VLT and near-IR  photometry images with the SOFI infrared array at the ESO NTT.
We have chosen to extend this sample by using the MACHO (Alcock et al.~2000b) 
and OGLE (Zebrun et al. 2001) databases to seven fields of the LMC bar,
at distances from 0.4 to 2.5 degrees away from the rotation center
(Soszynski et al. 2003). Two of these new fields are centered on the LMC
clusters NGC\,1835 and NGC\,2019.
We also observed additional stars in the fields LMC-4, LMC-9 and LMC-14 
defined in  Paper~I. As a control sample for  the 
LMC kinematic properties, we have selected
known long period variables (LPV) and Cepheids from MACHO and OGLE 
in the same fields (see also Paper~I). 
The observed fields are summarized in Table~\ref{Table:log} and shown in 
Fig.~\ref{Fig:centers} along with fields observed in Paper~I.
 
\begin{table*}[t]\tabcolsep=0.1pt\small
\begin{center}
\caption{The centers of the observed fields. The log of the observations.
}
\label{Table:log}
\begin{tabular}{l@{\hspace{5pt}}l@{\hspace{5pt}}l@{\hspace{7pt}}lccccccc}
\hline
\multicolumn{1}{l}{Event}{\hspace{5pt}}&
 \multicolumn{1}{c}{RA}&
\multicolumn{1}{c}{DEC}{\hspace{3pt}}&
\multicolumn{1}{c}{Instrument}{\hspace{3pt}}&
\multicolumn{1}{c}{Date of obs.}{\hspace{3pt}}&
\multicolumn{1}{c}{ExpTime (min)}{\hspace{3pt}}&
\multicolumn{1}{c}{No of Exp.}{\hspace{3pt}}&
\multicolumn{1}{c}{RRLyr's}{\hspace{3pt}}&
\multicolumn{1}{c}{Cep's}{\hspace{3pt}}&
\multicolumn{1}{c}{LPVs}{\hspace{3pt}}\\
\hline
LMC-4   &   05:17:14.6 & -70:46:59.0 &FORS2&28.12.2003&20&2&14&1&0 \\
LMC-9   &   05:20:20.3 & -69:15:12.0 &FORS2&28.12.2003&20&2&18&3&8\\
LMC-14  &   05:34:44.0 & -70:25:07.0 &FORS2&28.12.2003&20&2&14&0&3\\   
LMC-18  &   05:45:21.1 & -71:09:11.2 &FORS2&28.12.2003&20&2&2 &0&4\\
Ne01    &   05:18:01.0 & -69:29:53.0 &FORS2&28.12.2003&20&2&18&6&0\\
LMC-F1  &   05:28:28.4 & -69:33:48.8 &GMOS&17.03.2004 &30&2&8 &0&6\\
LMC-F2  &   05:37:19.2 & -70:00:39.0 &GMOS&20.03.2004 &30&2&6 &0&7\\
LMC-F4  &   05:05:35.0 & -68:44:45.7 &GMOS&22.03.2004 &30&2&7 &0&9\\
NGC1835 &   05:05:07.1 & -69:24:14.2 &GMOS&27.03.2004 &30&2&11&0&0\\ 
NGC2019 &   05:31:56.5&  -70:09:32.5 &GMOS&19.04.2004 &30&2&2 &0&13\\ 
\hline
\end{tabular}
\end{center}
\end{table*}

%
\begin{figure}[h]
\caption{MACHO image (http://www.macho.mcmaster. ca/)
  of LMC (R-band) showing
 the location of the MACHO fields and the fields for which 
 we obtained MOS spectra. 
 North is up and east is to the left. The 
 small thick 
 boxes indicate the  $7\arcmin \times\ 7\arcmin$ field of view of
 the FORS1/FORS2 CCD images and  $6\arcmin \times\ 6\arcmin$ 
 GMOS fields.
 }
 \label{Fig:centers}
\end{figure}

The whole sample (including data from Paper I) of 
RRab stars covers the period interval from
0.46 days to 0.78 days (only one star has a period of 0.97 days), 
with a mean value of $0.575\pm0.08$ days.
The RRc stars' periods range from 0.27 to 0.43 days
and the mean period is $0.36\pm0.04$ days. The RRe stars have
the mean period $0.28\pm0.01$ days.
These values are very similar to the values
determined by the MACHO team ${\langle P_{ab}\rangle=0.583}$ days (Alcock et al. 1996) 
from $\sim 7900$ RR Lyrae stars and the recent work of the
OGLE team (Soszynski et al. 2003): ${\langle P_{ab}\rangle=0.573}$
days. For RRc stars the mean 
periods are ${\langle P_c\rangle=0.342}$ days, from the MACHO team and 
${\langle P_c\rangle=0.339}$ from the OGLE team.

The new spectroscopic observations were taken with the FORS2 multi-slit 
spectrograph at the ESO Very Large Telescope (VLT) Unit Telescope 4 (UT4),
during the night of 28 December 2003, and with the
Gemini South multi object spectrograph (GMOS) in queue mode. 
We used FORS2 with the GRIS\_600B+22 grating that gives $R=780$ and
covers from $\lambda 3300$ to $\lambda 6210$ \AA.
GMOS was used with the B600+G5323 grating, which
is centered on $\lambda 5000$ \AA\, and gives $R=1688$.  
The resolution of FORS1 with GRIS\_600B+12 is also 780.
FORS2 has 19 movable slits, while in GMOS it is possible to put
a hundred slits in a single mask. The field of view of GMOS is
only 5.5 X 5.5 arcmin, while FORS2 has a field of view 6.8 X 6.8 arcmin.
The seeing during the FORS2 observations was 0.6-0.7 arcsec, while
GMOS spectra were taken with the mean seeing of 1 arcsec.
Both resolutions are adequate for the measurement of radial velocities
even in the broad-lined RR Lyrae star spectra, 
provided that a sufficient S/N is achieved.
During the pulsation of a RR Lyrae variable, the star's radial velocity
varies from the systemic velocity.
To avoid excessive broadening of spectral lines by changing velocity, 
the integration times of the 
spectroscopic observations should be kept under 5 percent of 
the pulsation cycle.
In  total, two exposures of 20 minutes with FORS2 and 30 minutes with 
GMOS were obtained  for each mask 
containing 5-10 RR Lyrae stars. We placed LPV and Cepheid variables on 
the remaining slits.
The FORS2 observations were taken in one night but are spaced in time 
to allow a better sampling of the 
radial velocity variation with phase. 
The GMOS observations are taken during 5 different nights.

In this extended sample we observed 100 RR Lyrae stars, 10 Cepheids, 
and 50 Miras with FORS2 and GMOS.

Using the same setup we also observed RR Lyrae stars of the globular 
cluster $\omega$ Cen with GMOS and repeated observations 
of some of the RR Lyrae stars in the LMC field, in 
order to assess the velocity errors.

The spectral data were 
reduced using  the standard  packages within IRAF.
HeNeAr lamps were used for the wavelength calibration.
Each spectrum was individually calibrated using an order 3 cubic
spline function for the fit and typically 14 usable lines.
The rms of the dispersion solution varies between 0.03 and 0.2 \AA~rms. 
The 0.2 \AA~rms of the wavelength calibration contributes about 12-13 km/sec
(at $\lambda \sim 5000$ \AA) to the error budget of the 
individual radial velocity
measurements and is added in quadrature to the other error sources.

To check the wavelength calibration, we measured several emission 
lines of the night sky that were also present in our 20 and 30 minute 
long exposures. The reference wavelengths of these lines were taken 
from the Keck telescope web page. In fifty six of our spectra we can 
measure the strong [OI] skyline at 5577.338\AA \,  and in some cases 
Na I at 5889.950 \AA. The central wavelengths of these lines were 
measured by fitting a Gaussian profile. The mean deviations of sky 
lines from their reference wavelengths varied from star to star and 
from $-1.18$~\AA\,  to $+0.74$\AA. Following Gallart et al. (2001) and Vivas 
et al (2005), these deviations were used to correct the zero point 
of the wavelength scale, which were applied to each spectrum by 
modifying the header parameter CRVAL1, the starting wavelength. The 
standard deviation of the mean offset of the sky lines from their 
reference wavelengths was typically of 0.09\AA\, (or $\sim 6$ km/s).

In summary, the daytime lamp spectra are used to determine the 
dispersion relation, but the zero point of the calibration, when 
possible is corrected by measuring the wavelengths of the skylines. 
Because we split the exposure time of each program star in two we 
extracted, wavelength calibrated and corrected each observation 
individually. Then, the two spectra were summed to improve the S/N. 
The final extracted and calibrated spectra have on average S/N = 30 
for FORS2 
and S/N = 25 for GMOS at 4750 \AA \, for LMC RR Lyrae stars. The S/N 
ratio for each star is given
in column 4 of Table~\ref{Table:RRLyrall}.

\section{Radial velocities}

The radial velocities of the stars were measured with the IRAF task 
fxcor, which performs a Fourier cross-correlation between a program star 
and a radial velocity standard star. The wavelength range for the 
cross-correlation was from 3800 to 5250 \AA, which includes hydrogen 
lines (H$\epsilon$, H$\beta$, H$\gamma$
and H$\delta$) and Ca II K $\lambda$ 3933.66 \AA. As expected with the 
high S/N of our spectra, the correlation peaks were well defined in all 
cases. However, they were broad because the correlations are dominated 
by the Balmer lines, which are wide features in these stars with A-F type 
spectra. 
Since the cross-correlation has to be made with a template that has a 
spectral type similar to the targets
and we did not observe any radial velocity standard during our runs,
we retrieved from the 
ESO archive the radial velocity standards
Kopff\,27(A3\,V), Feige\,56   (A0\,V) and HD\,155967(F6\,V). They were
taken with the same instrument and setup (FORS2, GRIS\_600B+22 grating, 
1 arcsec wide slit) under program 069.B-0343 (PI: Gallart). The spectra of 
these stars were reduced in the same way as the program stars, including 
a zero point correction with the skylines. After careful inspection we 
chose to use
as a template star HD\,155967, because its spectrum has the highest S/N. 
Each spectrum of RR Lyrae star was therefore correlated with the spectrum 
of this standard.

\subsection{Error analysis}
While fxcor returns a value for the standard deviation of the 
cross-correlation, this is not a true measure of the uncertainty 
because it does not take into account the zero-point uncertainty in 
the wavelength calibrations of the program star and the standard star. 
The sigmas of these zero-points were therefore added in quadrature to 
this value. For fifty six stars we added in quadrature the zero point 
corrections of the program stars plus the zero point correction of the 
HD\,155967 (0.02 \AA), and for the remaining stars, which have no measurable 
skylines in their spectra, we 
added the conservative estimate of 15 km/s. These final errors are 
listed in column 3 of Table~\ref{Table:RRLyrall}. 
Their mean value is 25$\pm1$  
km~s$^{-1}$.

To check the estimate of the observational errors and 
homogeneity of the sample
we compared the results obtained from multiple observations 
of some stars. There are 11 LMC RR Lyrae stars observed with  
FORS1 and FORS2 with four spectra. They are 
given in the first eleven rows of Table~\ref{Table:commonLMCRRLyr}.
The standard deviation of the measured velocities of these stars
is in the range from 2 to 35 km~s$^{-1}$, with a mean value of 18$\pm3$  
km~s$^{-1}$. The remaining thirty stars in the same table are RR Lyrae 
stars observed with GMOS on different nights, with 30 minute exposures, 
which have a high enough signal to noise ratio to 
measure the radial velocities on the individual, not summed 
spectra. The  mean value of the dispersion of these stars is 
25$\pm4$  km~s$^{-1}$. The whole sample of 41 stars
has a mean dispersion of 23$\pm3$  km~s$^{-1}$.
Table~\ref{Table:commonLMCRRLyr} 
lists the radial velocities of the common stars.

\begin{table*}[t]\tabcolsep=1pt\tiny
\begin{center}
\caption{LMC RR Lyrae common stars observed with FORS1, FORS2 and GMOS.}
\label{Table:commonLMCRRLyr}
\begin{tabular}{l@{ } l@{  }l@{   }l@{   }l@{   }c@{  }c@{   }c@{   }c@{   }}
\hline
\multicolumn{1}{c}{NAME}\hspace{0.1cm}&
\multicolumn{1}{c}{$<V>$}\hspace{0.5cm}&
\multicolumn{1}{c}{$RV_{1}$\hspace{0.5cm}}&
\multicolumn{1}{c}{$RV_{2}$\hspace{0.5cm}}&
\multicolumn{1}{c}{$RV_{MEAN}$\hspace{0.5cm}}&
\multicolumn{1}{c}{$\sigma_{RV}$\hspace{0.5cm}}\\
\hline
MACHO79.5508.735 \hspace{5cm}& 19.78\hspace{1cm} & 246\hspace{1cm} & 289\hspace{1cm} & 268 & 30 \\
MACHO13.5960.884 \hspace{1cm}& 19.23 & 226 & 223 & 225 & 2 \\
MACHO13.5961.623 \hspace{1cm}& 19.24 & 261 & 300 & 281 & 28 \\
MACHO13.5962.547 \hspace{1cm}& 19.21 & 283 & 275 & 279 & 6 \\
MACHO13.5962.656 \hspace{1cm}& 19.34 & 271 & 305 & 288 & 24 \\
MACHO80.6469.1657 \hspace{1cm}& 19.24 & 218 & 195 & 207 & 16 \\
OGLE052013.42-691153.0 \hspace{1cm}& 18.97 & 177 & 212 & 195 & 25 \\
OGLE052026.29-691815.0 \hspace{1cm}& 19.12 & 199 & 248 & 224 & 35 \\
MACHO11.8871.1122 \hspace{1cm}& 19.71 & 261 & 250 & 256 & 8 \\
MACHO11.8871.1299 \hspace{1cm}& 19.61 & 271 & 282 & 277 & 8 \\
MACHO11.8871.1447 \hspace{1cm}& 19.59 & 156 & 178 & 176 & 16 \\
OGLE052829.20-693201.1 \hspace{1cm}& 19.30 & 146 & 152 & 149 & 4 \\
OGLE052822.98-693310.6 \hspace{1cm}& 19.40 & 327 & 310 & 319 & 12 \\
OGLE052827.10-693347.7 \hspace{1cm}& 20.06 & 191 & 205 & 198 & 10 \\
OGLE052834.72-693326.4 \hspace{1cm}& 19.37 & 125 & 173 & 149 & 34 \\
OGLE052811.62-693423.8 \hspace{1cm}& 19.50 & 290 & 352 & 321 & 44 \\
OGLE052829.28-693442.0 \hspace{1cm}& 19.80 & 217 & 287 & 252 & 49 \\
OGLE052854.41-693403.8 \hspace{1cm}& 19.40 & 150 & 217 & 184 & 47 \\
OGLE052822.65-693529.9 \hspace{1cm}& 19.42 & 299 & 308 & 304 & 6 \\
OGLE053720.10-700202.5 \hspace{1cm}& 19.51 & 268 & 313 & 291 & 32 \\
OGLE053715.52-695812.2 \hspace{1cm}& 20.08 & 208 & 220 & 214 & 8 \\
OGLE053714.30-695846.0 \hspace{1cm}& 20.00 & 229 & 326 & 278 & 69 \\
OGLE053717.79-695923.9 \hspace{1cm}& 19.57 & 189 & 167 & 178 & 16 \\
OGLE053713.07-695956.8 \hspace{1cm}& 19.98 & 272 & 267 & 270 & 4 \\
OGLE053715.59-700105.8 \hspace{1cm}& 21.00 & 210 & 291 & 251 & 57 \\
OGLE050516.31-684514.9 \hspace{1cm}& 19.41 & 326 & 351 & 339 & 18 \\
OGLE050542.14-684529.5 \hspace{1cm}& 19.42 & 305 & 323 & 314 & 13 \\
OGLE05054491-6844330.0 \hspace{1cm}& 19.30 & 320 & 317 & 319 & 2 \\
OGLE050519.07-684522.5 \hspace{1cm}& 18.41 & 324 & 350 & 337 & 18 \\
OGLE050516.31-684514.9 \hspace{1cm}& 19.41 & 305 & 359 & 332 & 38 \\
OGLE050509.55-692505.5 \hspace{1cm}& 19.04 & 248 & 209 & 229 & 28 \\
OGLE050505.88-692500.9 \hspace{1cm}& 19.30 & 207 & 228 & 218 & 15 \\
OGLE050456.54-692449.0 \hspace{1cm}& 20.00 & 114 & 174 & 144 & 42 \\
OGLE050457.92-692419.2 \hspace{1cm}& 19.35 & 203 & 223 & 213 & 14 \\
OGLE050458.96-692447.2 \hspace{1cm}& 19.20 & 197 & 195 & 196 & 1 \\
OGLE050514.52-692412.2 \hspace{1cm}& 19.37 & 186 & 247 & 217 & 43 \\
OGLE050512.73-692446.5 \hspace{1cm}& 19.32 & 131 & 189 & 160 & 41 \\
OGLE050507.42-692346.6 \hspace{1cm}& 19.26 & 153 & 225 & 189 & 51 \\
OGLE050501.95-692343.5 \hspace{1cm}& 19.31 & 191 & 186 & 189 & 4 \\
OGLE050508.49-692330.7 \hspace{1cm}& 19.20 & 213 & 212 & 213 & 1 \\
OGLE050503.49-692319.2 \hspace{1cm}& 19.61 & 182 & 216 & 199 & 24 \\
\hline
\end{tabular}
\end{center}
\end{table*}

During the pulsation of a typical ab RR Lyrae variable, the star's radial 
velocity varies by $\pm$ 50-60 km/s about the systemic velocity (Layden 1994). 
In order to estimate the influence of the pulsation of the stars on the 
velocity dispersion 
we performed the following test. We selected eighteen RR Lyrae stars 
observed with GMOS with the best quality spectra and measured their radial 
velocities from the individual, not summed spectra. The OGLE and 
MACHO databases were used to derive the phases of the stars at the 
time of our observations. As in  Vivas et al.\ (2005) we fitted each star with 
the radial velocity curve of the well-studied RR Lyrae star X Arietis
to calculate the systemic velocity. We used Layden's (1994) 
parameterization of the velocity curve that Oke (1966) measured from the 
H$\gamma$ line. Following the discussion given in Vivas et al.\ (2005) and 
Layden (1994) the Balmer line curves are more appropriate for our purpose 
than the measurements based on the weak metal lines. The observations at 
phases less than 0.1 or greater than 0.85 were excluded from our test. In this
way we rejected only  one star which had phases 0.94 and 0.04. 
Then we calculated the systemic velocity at phase 0.50 for 
these stars. The mean value of these seventeen RR Lyrae stars calculated as an
average of the two measurements of the radial velocities and the mean value of
the systemic velocities do not differ much. The same is true for the respective
dispersions.  The mean value and dispersion of the 
averaged two measurements is 
RV=255$\pm65$ km~s$^{-1}$, while the mean and dispersion of the 
systemic velocities is RV=263$\pm63$ km~s$^{-1}$. The result is not 
surprising because for each star we have at least two observations, 
taken at random phase, which we averaged. The mean error between the average 
values of the individual radial velocities and systemic ones in our test 
sample is 9$\pm1$ km~s$^{-1}$. Since the mean radial velocity 
of the test sample and its dispersion are not very sensitive to the pulsation 
correction and  following the analysis from Paper I we decided to use radial 
velocities of RR
Lyrae stars uncorrected for pulsation.

Taking into account all the error estimates 
described above the upper
limit of accuracy of our radial velocity measurements is 25 km s$^{-1}$.
This value is in good agreement with the Gratton et al.\ (2004) 
estimates for 48 stars with multiple observations. 
We have to add in quadrature a conservative estimate of 9-10 km s$^{-1}$
from the phase correction, so the final velocity dispersion has 
to be corrected with   
${(\sigma_{rms}^2+\sigma_{phase}^2)}^{-1/2}$=27 km s$^{-1}$.

\subsection{Kinematics}

The data for the 
OGLE and MACHO LPVs and Cepheids 
are given in Table~\ref{Table:LPVs} and 
\ref{Table:Cepheids}, respectively. The properties of the
LMC RR Lyrae stars are summarized in Table~\ref{Table:RRLyrall}, 
where in columns 6, 7 and 8 
we list the mean V magnitude, period of the pulsation and type of star
taken from the MACHO and OGLE database. Radial velocities
are shown in column 2 and metallicities in column 5. 
These are discussed below.

\begin{table}\tabcolsep=1pt\tiny
\begin{center}
\caption{LPV stars from MACHO and OGLE databases observed with FORS2 and GMOS.}
\label{Table:LPVs}
\begin{tabular}{c@{ }c@{ }c@{ }c@{ }}
\hline
\multicolumn{1}{c}{NAME}&
\multicolumn{1}{c}{V}&
\multicolumn{1}{c}{$RV$ (km/s)}&
\multicolumn{1}{c}{Instrument}\\
\hline
OGLE05451038-7107080  & 14.97 & 225 & FORS2 \\
OGLE05452789-7106578  & 13.88 & 309 & FORS2 \\
OGLE05453011-7108405  & 16.21 & 187 & FORS2 \\
OGLE05452114-7110446  & 15.24 & 277 & FORS2 \\
OGLE05345471-7023475  & 14.65 & 217 & FORS2 \\
OGLE05344660-7025053  & 15.36 & 224 & FORS2 \\
OGLE05344085-7026341  & 19.2 & 249 & FORS2 \\
OGLE05202129-6912571  & 17.37 & 226 & FORS2 \\
OGLE05201078-6912417  & 14.99 & 235 & FORS2 \\
OGLE05202110-6914527  & 16.16 & 228 & FORS2 \\
OGLE05203074-6916455  & 19.85 & 193 & FORS2 \\
OGLE05201078-6912417  & 14.99 & 214 & FORS2 \\
OGLE05202948-6914132  & 17.64 & 199 & FORS2 \\
OGLE05201941-6914598  & 14.83 & 232 & FORS2 \\
OGLE05202789-6915096  & 15.88 & 227 & FORS2 \\
OGLE05283307-6931278 & 14.6 & 224 & GMOS \\
MACHO 77. 7795. 15   & - & 243 & GMOS \\
OGLE05280312-6932386 & 14.65 & 286 & GMOS \\
OGLE05285235-6933316 & 14.29 & 241 & GMOS \\
OGLE05373007-6958385 & 15.52 & 227 & GMOS \\
OGLE05372274-6958227 & 14.11 & 229 & GMOS \\
OGLE05370332-7000121 & 16.85 & 283 & GMOS \\
OGLE05370996-7000013 & 14.95 & 277 & GMOS \\
OGLE05365681-6959455 & 14.09 & 259 & GMOS \\
OGLE05373747-7001004 & 16.1 & 258 & GMOS \\
OGLE05370850-7002122 & 15.46 & 267 & GMOS \\
OGLE05372590-7002233 & 16.73 & 246 & GMOS \\
OGLE05371190-7002342 & 14.86 & 288 & GMOS \\
OGLE05060079-6843090 & 15.57 & 223 & GMOS \\
OGLE05055136-6843482 & 14.82 & 255 & GMOS \\
OGLE05054684-6843166 & 14.44 & 311 & GMOS \\
OGLE05054561-6842581 & 15.08 & 258 & GMOS \\
OGLE05055189-6844484 & 16.09 & 249 & GMOS \\
OGLE05050860-6844561 & 16.64 & 222 & GMOS \\
OGLE05060079-6845587 & 14.44 & 258 & GMOS \\
OGLE05055576-6846385 & 14.82 & 282 & GMOS \\
OGLE05054263-6846535 & 14.82 & 226 & GMOS \\
OGLE05322110-7008233  & 18.63 & 291 & GMOS \\
OGLE05314921-7008542  & 15.66 & 304 & GMOS \\
OGLE05320041-7008436  & 14.86 & 225 & GMOS \\
OGLE05301729-7007148  & 13.69 & 228 & GMOS \\
OGLE05300027-7008333 & 13.77 & 248 & GMOS \\
OGLE05314492-7008059  & 14.74 & 259 & GMOS \\
OGLE05320099-7009133  & 14.63 & 255 & GMOS \\
OGLE05315279-7009320  & 15.2 & 222 & GMOS \\
OGLE05315958-7010145  & 18.77 & 294 & GMOS \\
OGLE05315215-7010026  & 16.24 & 301 & GMOS \\
OGLE05321820-7009425  & 14.92 & 246 & GMOS \\
OGLE05321906-7009571  & 17.84 & 247 & GMOS \\
OGLE05315209-7011096  & 14.46 & 270 & GMOS \\    
\hline
\end{tabular}
\end{center}
\end{table}

\begin{table*}[t]\tabcolsep=0.1pt\tiny
\begin{center}
\caption{Cepheid stars from OGLE, MACHO and EROS database observed with FORS1 and FORS2}
\label{Table:Cepheids}
\begin{tabular}{l@{ } l@{ }  c@{ } c@{ } l@{} c@{} c@{}}
\hline
\multicolumn{1}{c}{Name}&
\multicolumn{1}{c}{RA}&
\multicolumn{1}{c}{DEC}&
\multicolumn{1}{c}{$RV$ (km/s)}&
\multicolumn{1}{c}{P(days)}&
\multicolumn{1}{c}{V(mag)}&
\multicolumn{1}{c}{Instrument}\\ 
\hline 
OGLE LMC-SC7 157350&05:18:33.63& -69:32:15.9& 270    &2.3425 & 15.841&FORS2  \\
EROS 2017&     05:17:40.78& -69:32:43.1& 241   &2.4813 &15.517&FORS2  \\
OGLE LMC-SC7 165501&05:18:28.17& -69:27:46.1& $310$   &4.3781 &15.819&FORS2  \\
EROS 2001&     05:17:50.90& -69:28:30.0& 263   &1.3087 &16.983&FORS2  \\
OGLE LMC-SC7 38692& 05:17:40.78& -69:32:43.1& 211   &2.4813 &15.517&FORS2  \\
OGLE LMC-SC6 118148&05:20:21.00& -69:12:21.0& 259   &10.511 &17.072&FORS2  \\
OGLE LMC-SC6 102475&05:20:04.63& -69:16:50.9& 236  &3.3589 &15.817&FORS2   \\
MACHO 11.8870.40&  05:34:58.00& -70:26:34.8& 232   & 2.9827 &16.264&FORS1\\
MACHO 11.8622.24&  05:33:58.00& -70:52:17.4& 261  & 2.0921 & 15.901&FORS1\\  
MACHO 80.6468.46&  05:20:04.00& -69:16:51.2& 272    & 3.3587 & 15.821&FORS1\\
MACHO 80.6348.23&  05:19:47.00& -69:12:29.8& 248   & 4.7348 & 15.695&FORS1\\  
MACHO 80.6468.21&  05:19:54.00& -69:15:27.4& 281   & 3.1557 & 15.523&FORS1\\  
\hline
\end{tabular}
\end{center}
\end{table*}

In summary, the whole dataset (FORS1, FORS2 and GMOS) contains 
26 RR Lyrae stars in $\omega$ Cen, 73 LPV 
variable stars, 12 Cepheids, 137 RR Lyrae stars in the field of LMC and 13
RR Lyrae stars in the LMC globular clusters NGC\,1835 and NGC\,2019.

Following the method described in Minniti et al.\ (2003) we calculated
the dispersions for the velocity measurements for the field RR Lyrae stars,
cluster RR Lyrae stars, LPVs and Cepheid stars. They are given in Table~\ref{Table:observedRVsigma}.
The true velocity dispersions of LMC Cepheid and LPV stars are taken from the literature (see Paper~I
for more details.)

Since we measured $\sigma_{obs}=57\pm3$ km~s$^{-1}$ 
for the LMC field RR Lyrae stars and assuming 
${(\sigma_{rms}^2+\sigma_{phase}^2)}^{-1/2}$=27 km~s$^{-1}$, we obtain
$\sigma_{true}= 50\pm2$ km s$^{-1}$. 
Here $\sigma_{obs}$ is the observed velocity dispersion, calculated as 
the standard deviation of the individual velocities from the average value,
$\sigma_{rms}$ is the mean error of the individual velocity measurements and
$\sigma_{phase}$ is the mean dispersion in the velocities because we observed
the stars at a random phase. The errors of the observed and true velocity dispersions were 
calculated as a standard deviation of the mean values divided by the 
square root of the number of the stars. 
Thus we confirm our 
result from Minniti et al.\ (2003), 
that a kinematically hot metal-poor old halo exists in the LMC.
The average (mean) radial velocity (RV) derived for the LMC field RR Lyrae stars is RV=258$\pm$5 
km~s$^{-1}$, in agreement with the Gratton et al. (2004)  value of 
RV=261 km~s$^{-1}$ and with the value of 257 km~s$^{-1}$
found for LMC red giants by Cole et al. (2005).  

The $\sigma_{obs}$ for RR Lyrae stars of the globular clusters 
NGC\,1835 and NGC\,2019 is 28 km s$^{-1}$, and 
$\sigma_{true}= 7\pm2$ km~s$^{-1}$. Freeman et al. (1983) have shown that the clusters older than 
about 0.3 Gyr appear to lie in a flat rotating disk whose velocity 
dispersion is only 15 km~s$^{-1}$. 
This result for the clusters was confirmed by 
Schommer et al. (1992). 

We measured only 13 RR Lyrae stars in NGC\,1835 and NGC\,2019 and find that
the true velocity dispersion of the cluster RR Lyrae is
much smaller than the velocity dispersion of the field RR Lyrae stars.
Dubath at al. (1997) measured the 
core velocity dispersions for 10 old Magellanic
globular clusters from integrated-light spectra. For NGC\,1835 
they reported 10.4 km/s and 7.5 km/s for NGC\,2019. The velocity dispersion
obtained from our 11 RR Lyrae stars of NGC\,1835 is
$\sigma_{true}= 7$ km s$^{-1}$. 
For NGC\,2019 we have only 2 stars and it is not 
possible to calculate a realistic value.
The average radial velocity derived for the NGC\,1835 RR Lyrae stars 
is RV=198$\pm$ 6 km s$^{-1}$,
while for the  NGC\,2019 we obtained 231 $\pm$ 2  km s$^{-1}$ from two
stars, again in agreement with the average value measured for these clusters by Freeman et al. (1983)
and Schommer et al. (1992).

For the 73 LPV stars in the LMC we obtained a mean radial velocity 
RV=252$\pm5$ km s$^{-1}$ and observed velocity dispersion of 35$\pm3$ km s$^{-1}$.

\begin{table*}[t]\tabcolsep=0.1pt\small
\begin{center}
\caption{Observed velocity dispersions}
\label{Table:observedRVsigma}
\begin{tabular}{l@{ }l@{ }l@{ }l@{}l@{}l@{}}
\hline
\multicolumn{1}{c}{Population}\hspace{0.5cm}&
\multicolumn{1}{c}{Number of stars}\hspace{0.5cm}&
\multicolumn{1}{c}{$RV_{mean}$}\hspace{0.5cm}&
\multicolumn{1}{c}{$\sigma_{obs}$}\hspace{0.5cm}&
\multicolumn{1}{c}{${(\sigma_{rms}^2+\sigma_{phase}^2)}^{-1/2}$}\hspace{0.5cm}& 
\multicolumn{1}{c}{$\sigma_{true}$}\\
\hline
LMC Cepheid stars          &\hspace{0.5cm}  12&  262$\pm8$&   22$\pm5$&\hspace{0.7cm}  16$\pm3$  &   15\\
LMC LPV stars              &\hspace{0.5cm}  73&  252$\pm5$&   35$\pm3$&\hspace{0.7cm}  12$\pm2$  &   33\\
LMC RR Lyrae field stars   &\hspace{0.5cm} 137&  258$\pm5$&   57$\pm3$&\hspace{0.7cm}  27  &   50$\pm2$\\
LMC RR Lyrae cluster stars &\hspace{0.5cm}  13&  203$\pm6$&   28$\pm6$&\hspace{0.7cm}  27  &   7$\pm2$\\
\hline
\end{tabular}
\end{center}
\end{table*}

To test the accuracy and stability of the velocity dispersion measurements we
calculated the observed values at different radii with respect to the
rotation center given in Soszynski et al. (2003). 
The results are 
given in Table~\ref{Table:meanRV_soszynsky}. As can be seen the dispersion is 
relatively stable and does not depend on the radius up to the
limit of 2.0 degree. We excluded the last bin (2.0-2.5 arcmin) because it contains only 2 
objects. Their RV and sigma differ from the values in the other bins, 
but the sample is too small. 

\begin{table}[t]\tabcolsep=0.1pt\small
\begin{center}
\caption{Mean radial velocity and dispersion as a function of distance 
from the center of rotation (Soszynski et al.\ 2003) for RR Lyrae 
field stars in the LMC.}
\label{Table:meanRV_soszynsky}
\begin{tabular}{c@{\hspace{0.5cm} }c@{\hspace{0.5cm} }c@{\hspace{0.5cm} }l@{\hspace{0.5cm} }}
\hline
\multicolumn{1}{c}{Radius\hspace{0.5cm}}&
\multicolumn{1}{c}{$N$\hspace{0.5cm}}&
\multicolumn{1}{c}{$RV$\hspace{0.5cm}}&
\multicolumn{1}{l}{$\sigma_{RV}$\hspace{1cm}}\\
\hline
0.0-0.5 &43 & 257 & 61$\pm7$ \\
0.5-1.0 &12 & 247 & 76$\pm15$ \\
1.0-1.5 &72 & 255 & 49$\pm4$\\
1.5-2.0 &9  & 225  & 54$\pm9$\\
2.0-2.5 &2 & 316 & 19$\pm10$\\
\hline
\end{tabular}
\end{center}
\end{table}

Another test of the dispersion stability is to check the influence of
possible outliers. Since the mean V magnitude taken from OGLE and 
MACHO databases 
ranges from 18.33 to 21.0 (with the average luminosity of the LMC
RR Lyrae stars being around V=19.4 mag)  the brightest/faintest stars 
could either not be RR Lyraes or not belong to the LMC.
Thus we removed the brightest five and the faintest nine stars and 
calculated the mean velocity and dispersion using only RR Lyrae stars 
in the magnitude interval from 18.89 to 19.91. The mean radial velocity 
and its dispersion is practically the same 
as for the whole sample: 257 and 57 km s$^{-1}$, respectively. 
The velocity dispersion is slightly smaller (55 km s$^{-1}$),
if we take into consideration only 56 stars for which we correct 
the zeropoint with the skylines (see Section 2).

\begin{figure}[h]
\resizebox{\hsize}{!}{\includegraphics{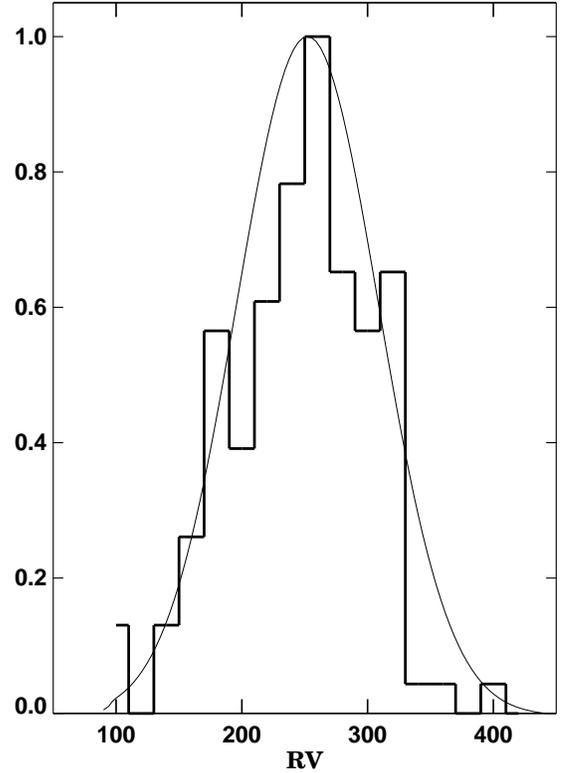}}
 \caption{The radial velocity distributions of RR Lyrae field 
 stars.
 The solid line represents the best Gaussian fit.
 }
\label{Fig:RVbinned_gauss}
\end{figure}

To test if the radial velocity distribution 
is Gaussian or shows evidence of multiple components we 
plotted in Fig.~\ref{Fig:RVbinned_gauss} the histogram of the radial velocities of RR
Lyrae field stars binned to 20 km s$^{-1}$
and the best Gaussian fit to the data.
The Kolmogorov-Smirnov test finds that the LMC RR Lyrae star distribution is consistent
with a normal distribution with a probability of 99.9$\%$
and we have not found evidence for different components.

 %

\begin{figure}[h]
\resizebox{\hsize}{!}{\includegraphics{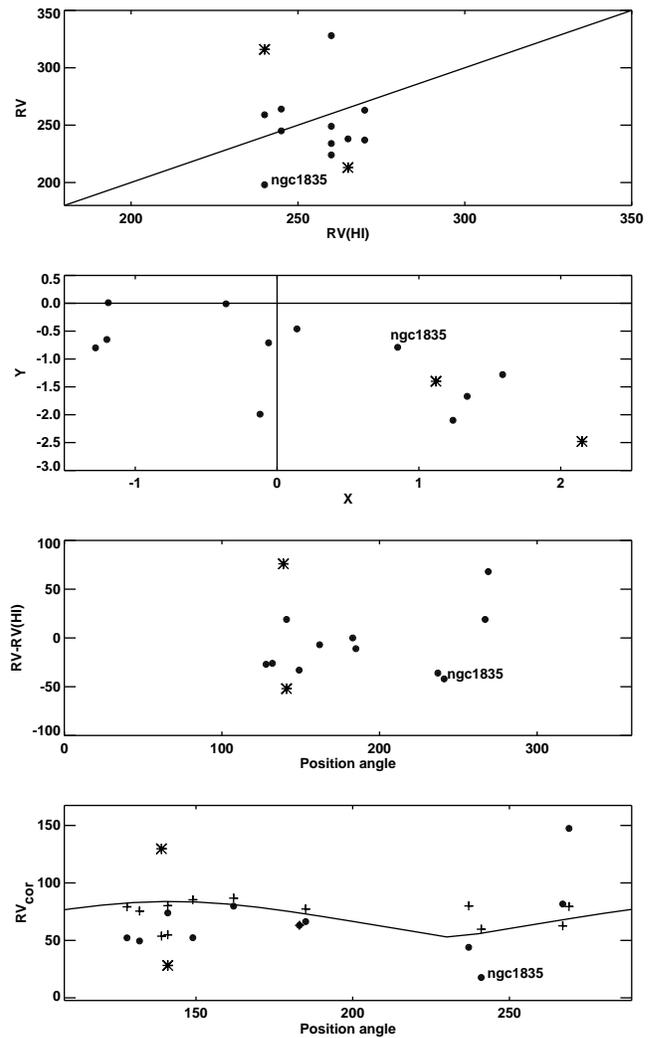}}
 \caption{Top panel: The mean radial velocity (RV) of the observed RR Lyrae field 
 stars vs. radial velocity of HI. The mean radial velocity of the RR Lyrae 
 globular cluster NGC\,1835 stars is labeled.  The
 asterisks mark the fields LMC-18 and NGC\,2019, which are excluded from the analysis 
 (see the text). The solid line is the line of equality. 
 Second panel: The rectangular X, Y coordinates in degrees of the observed 
 fields.
 Third panel: The position angle vs. differences between the radial 
 velocities of the observed fields and HI radial velocities.
 Bottom panel: The position angle vs. galactocentric velocities of RR Lyrae stars 
 (points) and HI (crosses).
 The best fit rotation  solution for HI velocities is shown as a solid line.}
\label{Fig:PA_velocities}
\end{figure}

To measure the rotation we follow the method described in Schommer et al (1992),
Freeman et al.\ (1983) and Feitzinger \& Weiss (1979). 
First we measured the radial velocities of HI regions using the HI survey 
performed with the Parkes telescope in Australia (Bruns et al.\ 2005) 
at the positions of the centers of our fields. They are given in Column 3 of
Table~\ref{Table:positions_velocities}.
The observed mean velocities of the RR Lyrae star population in each of the fields 
is compared with the HI velocity in the same field in the top panel of 
Fig.~\ref{Fig:PA_velocities}. The mean radial velocity of the globular cluster
NGC\,1835 is labeled. The fields LMC-18 and NGC\,2019 are excluded 
from the analysis, because they have only two RR Lyrae stars. They are 
marked with asterisks. Another three fields LMC-F4 (7 RR Lyrae stars), 
LMC-7 (5 RR Lyraes) and LMC-F2 (6 RR Lyraes) also have very poor statistics.
Equality in  Fig.~\ref{Fig:PA_velocities} is indicated by the line of slope unity. 
From the remaining seven fields, two have radial velocities of RR Lyrae 
stars higher than RV(HI), one with the same RV, and four with RV of the 
RR Lyrae stars lower than RV(HI).
Freeman et al. (1983) compared the HI velocities
and the observed velocities of 35 globular clusters in the LMC.
They found that the young and intermediate age globular clusters have the same
velocity distribution as the gas, but the older clusters in the LMC
have significant systematic offset to lower velocities. We also can see 
some weak tendency to lower mean velocities of RR Lyrae stars than
HI ones, but our data are not enough to outline it.
 
We transformed equatorial coordinates RA and DEC of our fields to 
the rectangular X,Y system using equation 1 of Feitzinger \& Weiss (1979)
with the origin of the coordinate system the center of the LMC as 
defined by the HI map ($RA=5^h20\fm39.84$, $DEC=-69^\circ 14'9.5$). 
This is shown in the second panel
of Fig.~\ref{Fig:PA_velocities}. The position angle (PA) was calculated 
within this
coordinate system, and it increases from
N through E. The differences between the radial velocities of the 
observed fields and HI radial velocities do not show a trend with 
position angle. We transformed the observed heliocentric 
radial velocities of the RR Lyrae populations in our fields and of HI 
to the galactocentric system,
using 220 km/s (van der Marel et al. 2002) for the Galactic rotation. 
These radial velocities are given in Columns 7 and 8 of
Table~\ref{Table:positions_velocities}. 
To study the rotational properties of RR Lyrae stars we fit the equation from
Schommer at al (1992):
\begin{eqnarray*}
  V(\theta)&=& V_m{[tan(\theta-\theta_0)\times sec(i)]^2+1}^{-0.5}+V_{sys}\hspace{10pt} \\
 \end{eqnarray*}
where $V_m$ is the amplitude of the rotation velocity, $\theta_0$ is the 
orientation of the line of nodes, and $V_{sys}$ is the systemic velocity,
while $i$ is the inclination angle, equal to 34.7 (van der Marel et al. 2002).
The best fit rotation  solution for HI velocities is obtained for  
$V_m=32$, $V_{sys}=49$ and $\theta_0=172$.
For RR Lyrae velocities it was not possible to obtain an adequate fit.
One reason could be that the mean velocities
are less precise than the velocity dispersion or that 
our fields extend only over the position angles
from 120 to 260 degrees.

%

\begin{table*}[t]\tabcolsep=0.1pt\small
\begin{center}
\caption{Positions and velocities for observed fields in the LMC}
\label{Table:positions_velocities}
\begin{tabular}{l@{\hspace{0.5cm} }c@{\hspace{0.5cm} }c@{\hspace{0.5cm} }c@{\hspace{0.5cm} }r@{\hspace{0.5cm} }r@{\hspace{0.5cm}}r@{\hspace{0.5cm}}r@{\hspace{0.5cm}}c@{}}
\hline
\multicolumn{1}{c}{Field\hspace{0.5cm}}&
\multicolumn{1}{c}{$RV$\hspace{0.5cm}}&
\multicolumn{1}{c}{$RV(HI)$\hspace{0.5cm}}&
\multicolumn{1}{c}{Radius\hspace{0.7cm}}&
\multicolumn{1}{c}{X\hspace{0.7cm}}&
\multicolumn{1}{c}{Y\hspace{0.7cm}}&
\multicolumn{1}{c}{$RV_{cor}$\hspace{0.7cm}}& 
\multicolumn{1}{c}{$RV(HI)_{cor}$\hspace{0.7cm}}& 
\multicolumn{1}{c}{$PA$\hspace{0.7cm}}\\
\hline
LMC-1  &  264 &245&    0.69&  -0.36 & -0.01 & 81.65 & 62.65  &  268 \\
LMC-4  &  245 &245&    1.57&  -0.12 & -1.99 & 63.05 & 63.05  &  184 \\
LMC-7  &  224 &260&    1.49&  -1.28 & -0.80 & 43.95 & 79.95  &  238 \\
LMC-9  &  263 &270&    0.03&  0.14 & -0.46 & 79.78 & 86.78  &   163 \\
LMC-12 &  237 &270&    1.94&  1.24 & -2.10 & 52.30 & 85.30  &   149 \\
LMC-14 &  259 &240&    1.67&  1.34 & -1.67 & 73.88 & 54.88  &   141 \\
LMC-18 &  316 &240&    2.77&  2.15  & -2.48 & 129.69 & 53.69  &   139 \\
Ne01  &   249 &260&    0.35&  -0.06 & -0.71 & 66.32 & 77.32  &  185 \\
LMC-F1 &  234 &260&    0.76&  0.85 & -0.79 & 49.46& 75.46  &   133 \\
LMC-F2 &  238 &265&    1.62&  1.59 & -1.28 & 52.20 & 79.20  &   129 \\
LMC-F4 &  328 &260&    1.45&  -1.19 & 0.01 & 147.41 & 79.41  &   269 \\
NGC1835&  198 &240&    1.38&  -1.20  & -0.65 &17.70  & 59.70  & 242\\
NGC2019&  231 &265&    1.33&    1.12  & -1.40  &  28.19&  80.19 & 141\\
\hline
\end{tabular}
\end{center}
\end{table*}

\section{Metallicities}

The metallicities are measured using the Gratton et al. (2004) method, which 
calibrate the so called metallicity index using spectra of variable and 
constant HB stars of two globular 
clusters, M\,68 and NGC\,1851. We used the metallicity calibrations
described by equations 3,4,5 and 6 of Gratton et al. (2004) 
and the derived values are listed 
in Column 5 of Table~\ref{Table:RRLyrall}.  
We estimate a global uncertainty of 0.20 dex$\pm0.02$ in the metallicity 
as a mean value of
18 stars observed with GMOS for which we measured the 
metallicities on the 
individual, not summed spectra. 
The error of this value was calculated as the standard deviation of the
mean uncertainty divided by the square root of the number of the stars.

The average [Fe/H] value of our 104 field RR Lyrae stars is  
$\mathrm{[Fe/H]} =-1.53\pm0.02$ dex, which is in good agreement with 
the Clementini et al.\ (2003) and Gratton at al.\ (2004) 
average value of [Fe/H] =$-1.48\pm0.03$ dex. These metallicities are on  
the Harris (1996) metallicity scale and are on average 0.06 dex more 
metal-rich than the Zinn and West (1984) scale (Gratton at al. 2004).

The mean metallicity of 5 RRe stars is [Fe/H] =$-1.52\pm0.09$ dex, 26
RRc stars have [Fe/H] =$-1.53\pm0.02$ dex and 73  RRab stars have
[Fe/H] =$-1.54\pm0.03$ dex. Therefore we confirm that 
there are no metallicity differences among Bailey's types (Gratton et al.\ 2004).
The average metallicity of NGC\,1835 is [Fe/H] =$-1.98\pm0.06$ dex.
Thus, the field RR Lyrae stars seem to be more metal rich than 
globular cluster RR Lyrae stars.

\begin{figure}
\resizebox{\hsize}{!}{\includegraphics{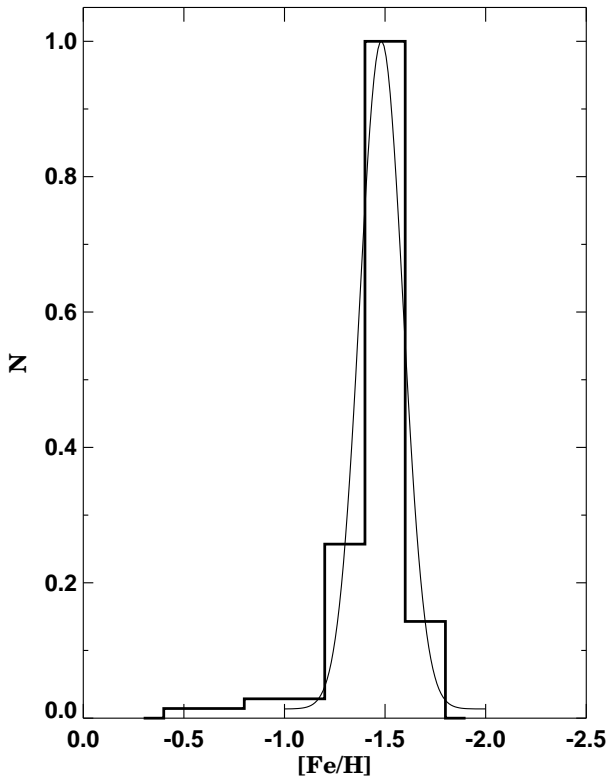}}
 \caption{The metallicity histogram for RR Lyrae stars in our fields.
The solid line represents the Gaussian distribution.
}
 \label{Fig:FeHgauss}
\end{figure}

The metallicity distribution of all the field RR Lyrae stars 
is shown in Fig.~\ref{Fig:FeHgauss}
and is well fitted by the Gaussian distribution centered on 
[Fe/H] =$-1.48$. 

%

In Table~\ref{Table:RV_FeH} we report the  mean velocity (RV) 
and dispersion ($\sigma_{RV}$) of different metallicity bins. 
The dispersion seems to be stable and does not depend on the metallicity.
The first metallicity bin shows lower mean radial velocity and dispersion, but 
the sample is very small - this bin contains only 3 stars. 
This is in contrast to the result of Cole et al.\ (2005), who found
that the most metal poor RGB stars show higher velocity dispersion. However,
they investigated different stellar populations, while in our work we have
analyzed the RR Lyrae population in different metallicity bins.

\begin{table}[t]\tabcolsep=0.1pt\small
\begin{center}
\caption{Mean radial velocity and dispersion for different metallicity 
bins for field RR Lyrae stars in the LMC}
\label{Table:RV_FeH}
\begin{tabular}{c@{\hspace{0.5cm} }c@{\hspace{0.5cm} }c@{\hspace{0.5cm} }c@{\hspace{0.5cm} }}
\hline
\multicolumn{1}{c}{[Fe/H]\hspace{0.5cm}}&
\multicolumn{1}{c}{$N$\hspace{0.5cm}}&
\multicolumn{1}{c}{$RV$\hspace{0.5cm}}&
\multicolumn{1}{c}{$\sigma_{RV}$\hspace{1cm}}\\
\hline
-0.5$-$-1.0 &3 & 199 & 30$\pm17$ \\
-1.0$-$-1.5 &21 & 248 & 51$\pm11$\\
-1.5$-$-2.0 &80  &  260  & 57$\pm6$\\
\hline
\end{tabular}
\end{center}
\end{table}

\section{Summary and conclusions}
We have presented new low-resolution spectra for 100 RR Lyrae field stars in the 
Large Magellanic Cloud. This sample more than triples that presented in 
Paper~I and allows us to strengthen our previous results regarding the kinematics 
and metallicity properties of this old and metal-poor population
in our nearest neighboring galaxy.

The average metallicity of the whole sample of RR Lyrae stars in the LMC is 
[Fe/H]= -1.53$ \pm 0.02$ dex.
The metallicity distribution is well fitted by a single Gaussian.
There is no significant variation of the velocity dispersion with metallicity. 
This is at odds with the results of Cole et al.\ (2005) who found an
increasing velocity dispersion with decreasing metallicity. However, these authors looked at 
different stellar populations in the LMC, while our sample comprises only RR Lyrae stars. 
Thus the trend between the metallicity and velocity dispersion may be due to age rather 
than to metallicity.

Our main result is that the mean velocity dispersion of the LMC field RR Lyrae 
stars is $\sigma_{RV} = 50 \pm 2$ km/s. This quantity does not appear 
to vary with distance from the LMC center up to 2 degrees. 

The mean velocities show some field to field variation, and
our conclusions are weaker for this quantity. They appear to differ
slightly from the velocities of the HI gas, but our data do not allow us to 
measure the bulk rotation.

\begin{acknowledgements}
DM is supported by Fondap Center for Astrophysics 15010003 
and by a Fellowship from the John Simon Guggenheim Foundation.
We thank our referee, Dr. Gisella Clementini, for very useful comments 
which helped to improve this manuscript significantly.
\end{acknowledgements}

\clearpage
\begin{longtable}{lccccccl}
\caption{\label{Table:RRLyrall} The whole sample of the LMC RR Lyrae stars taken with 
FORS1, FORS2 and GMOS.}\\
\hline\hline
name&  $RV$ (km/s)& $\sigma(RV)$ & $S/N$ & [Fe/H]& V & Period(d)&Type\\
\hline 
\endfirsthead
\caption{continued.}\\
\hline\hline
name&  $RV$ (km/s)& $\sigma(RV)$ & $S/N$ & [Fe/H]& V & Period(d)&Type\\
\hline
\endhead
\hline
\endfoot
 MACHO10.3802.311 & 234 & 20 & 38 & -1.51 & 18.99 & 0.5460 & RRab \\
MACHO10.3802.339 & 243 & 25 & 25 & -1.64 & 19.37 & 0.5120 & RRab \\
MACHO10.3802.446 & 165 & 21 & 17 & -1.60 & 19.49 & 0.3080 & RRc \\
MACHO10.3922.978 & 273 & 31 & 27 & -1.87 & 19.09 & 0.5568 & RRab \\
MACHO10.3923.351 & 204 & 40 & 22 & -1.60 & 19.24 & 0.3339 & RRc \\
MACHO11.8622.757 & 263 & 33 & 26 & -1.59 & 19.11 & 0.6860 & RRab \\
MACHO11.8623.3792 & 221 & 12 & 32 & -1.49 & 19.31 & 0.6149 & RRab \\
MACHO11.8623.779 & 171 & 23 & 32 & -1.54 & 19.20 & 0.6150 & RRab \\
MACHO11.8623.826 & 290 & 14 & 21 & -1.57 & 18.95 & 0.5920 & RRab \\
MACHO11.8744.658 & 197 & 18 & 37 & -1.52 & 18.92 & 0.4155 & RRc \\
MACHO11.8744.752 & 195 & 8 & 29 & -1.56 & 19.14 & 0.2809 & RRe \\
MACHO11.8744.830 & 325 & 23 & 30 & -1.53 & 19.47 & 0.5510 & RRab \\
MACHO11.8749.1208 & 269 & 27 & 28 & -1.56 & 19.56 & 0.4757 & RRab \\
MACHO11.8749.1324 & 116 & 38 & 32 & - & 19.63 & 0.5120 & RRab \\
MACHO11.8750.1425 & 261 & 26 & 38 & -1.52 & 19.26 & 0.3487 & RRc \\
MACHO11.8750.1672 & 208 & 24 & 24 & -1.51 & 19.62 & 0.3400 & RRc \\
MACHO11.8750.1827 & 252 & 16 & 35 & -1.60 & 19.74 & 0.5186 & RRab \\
MACHO11.8750.2045 & 299 & 19 & 32 & -1.82 & 19.81 & 0.4763 & RRab \\
MACHO11.8870.1275 & 274 & 16 & 37 & -1.60 & 19.33 & 0.4638 & RRab \\
MACHO11.8871.1096 & 308 & 11 & 23 & - & 19.43 & 0.5473 & RRab \\
MACHO11.8871.1122 & 256 & 32 & 24 & -1.49 & 19.71 & 0.5014 & RRab \\
MACHO11.8871.1299 & 277 & 28 & 26 & -1.49 & 19.61 & 0.5957 & RRab \\
MACHO11.8871.1362 & 250 & 17 & 30 & -1.58 & 19.61 & 0.6064 & RRab \\
MACHO11.8871.1447 & 176 & 16 & 28 & -1.50 & 19.59 & 0.2640 & RRe \\
MACHO11.8871.1516 & 262 & 15 & 27 & - & 19.46 & 0.5463 & RRab \\
MACHO13.5839.1023 & 188 & 38 & 25 & -0.75 & 19.66 & 0.5823 & RRab \\
MACHO13.5840.608 & 252 & 30 & 33 & -1.53 & 19.32 & 0.5188 & RRab \\
MACHO13.5840.730 & 258 & 31 & 31 & -1.79 & 19.41 & 0.6215 & RRab \\
MACHO13.5840.768 & 231 & 25 & 28 & -1.63 & 19.65 & 0.5515 & RRab \\
MACHO13.5960.884 & 225 & 21 & 28 & -1.56 & 19.23 & 0.3459 & RRc \\
MACHO13.5961.435 & 319 & 18 & 36 & -1.51 & 18.84 & 0.5088 & RRab \\
MACHO13.5961.511 & 200 & 26 & 22 & -1.69 & 19.1 & 0.7100 & RRab \\
MACHO13.5961.623 & 281 & 22 & 26 & -1.60 & 19.24 & 0.6180 & RRab \\
MACHO13.5961.648 & 293 & 24 & 30 & - & 19.39 & 0.5670 & RRab \\
MACHO13.5961.720 & 233 & 24 & 22 & -0.92 & 19.42 & 0.6090 & RRab \\
MACHO13.5962.547 & 279 & 28 & 36 & -1.71 & 19.21 & 0.6464 & RRab \\
MACHO13.5962.656 & 288 & 22 & 26 & -1.65 & 19.34 & 0.6090 & RRab \\
MACHO13.6082.701 & 175 & 18 & 24 & - & 19.41 & 0.5525 & RRab \\
MACHO13.6082.742 & 215 & 24 & 30 & - & 19.33 & 0.5163 & RRab \\
MACHO2.5507.5945 & 315 & 19 & 28 & - & 19.54 & 0.6234 & RRab \\
MACHO2.5507.6046 & 338 & 22 & 25 & - & 19.63 & 0.4920 & RRab \\
MACHO2.5508.3096 & 196 & 32 & 9 & - & 19.96 & 0.5593 & RRab \\
MACHO2.5628.5690 & 225 & 24 & 13 & - & 19.69 & 0.6156 & RRab \\
MACHO2.5628.6276 & 177 & 38 & 5 & -0.57 & 19.91 & 0.4781 & RRab \\
MACHO79.5507.1039 & 317 & 31 & 7 & -1.32 & 19.59 & 0.5608 & RRab \\
MACHO79.5507.1039 & 325 & 27 & 7 & - & 19.59 & 0.5608 & RRab \\
MACHO79.5507.1485 & 313 & 22 & 24 & -1.60 & 19.31 & 0.6234 & RRab \\
MACHO79.5507.1580 & 244 & 35 & 11 & - & 19.46 & 0.9720 & RRab \\
MACHO79.5508.427 & 279 & 28 & 15 & -1.66 & 19.73 & 0.5593 & RRab \\
MACHO79.5508.534 & 339 & 10 & 31 & -1.62 & 19.52 & 0.5723 & RRab \\
MACHO79.5508.682 & 216 & 27 & 17 & -1.77 & 19.89 & 0.7591 & RRab \\
MACHO79.5508.735 & 268 & 14 & 33 & -1.54 & 19.78 & 0.5057 & RRab \\
MACHO79.5628.1065 & 236 & 34 & 7 & -1.24 & 19.84 & 0.5305 & RRab \\
MACHO79.5628.1300 & 248 & 35 & 15 & - & 19.42 & 0.6160 & RRab \\
MACHO79.5628.1597 & 274 & 32 & 5 & - & 19.51 & 0.2812 & RRe \\
MACHO79.5628.1650 & 187 & 19 & 20 & - & 19.26 & 0.3389 & RRc \\
MACHO79.5628.2110 & 259 & 35 & 10 & - & 19.8 & 0.5260 & RRab \\
MACHO80.6347.1940 & 290 & 30 & 24 & -1.09 & 19.21 & 0.4083 & RRc \\
MACHO80.6467.2128 & 295 & 25 & 12 & - & 19.35 & 0.5805 & RRab \\
MACHO80.6468.1883 & 277 & 18 & 46 & -1.51 & 18.92 & 0.3346 & RRc \\
MACHO80.6468.2616 & 268 & 23 & 42 & -1.64 & 19.21 & 0.3684 & RRc \\
MACHO80.6468.2799 & 207 & 23 & 25 & -1.79 & 19.32 & 0.5731 & RRab \\
MACHO80.6469.1657 & 207 & 14 & 29 & -1.61 & 19.24 & 0.5804 & RRab \\
MACHO80.6469.1712 & 321 & 14 & 47 & -1.53 & 18.84 & 0.7720 & RRab \\
MACHO80.6588.1605 & 294 & 23 & 35 & - & 18.96 & 0.5237 & RRab \\
MACHO80.6588.2703 & 353 & 26 & 28 & -1.60 & 19.23 & 0.4794 & RRab \\
MACHO80.6589.2425 & 278 & 23 & 36 & -1.62 & 19.51 & 0.3497 & RRc \\
OGLE050456.54-692449.0 & 144 & 39 & 5 & -2.15 & 20.00 & 0.2779 & RRe \\
OGLE050457.92-692419.2 & 213 & 31 & 16 & -2.31 & 19.35 & 0.6000 & RRab \\
OGLE050458.96-692447.2 & 196 & 31 & 15 & -2.05 & 19.20 & 0.5157 & RRab \\
OGLE050501.95-69:23:43.5 & 189 & 33 & 15 & -1.97 & 19.31 & 0.3216 & RRc \\
OGLE050503.49-692319.2 & 199 & 30 & 16 & -2.22 & 19.61 & 0.3350 & RRc \\
OGLE050505.88-692500.9 & 218 & 25 & 27 & -1.94 & 19.30 & 0.5462 & RRab \\
OGLE050507.42-692346.6 & 189 & 28 & 22 & -1.85 & 19.26 & 0.5382 & RRab \\
OGLE050508.49-692330.7 & 213 & 28 & 17 & -1.66 & 19.20 & 0.3674 & RRc \\
OGLE050509.55-692505.5 & 244 & 24 & 22 & -1.9 & 19.04 & 0.3591 & RRc \\
OGLE05051028-6845491 & 350 & 22 & 12 & - & 20.66 & 0.5234 & RRab \\
OGLE050512.73-692446.5 & 160 & 32 & 15 & -2.01 & 19.32 & 0.3785 & RRc \\
OGLE050514.52-692412.2 & 217 & 33 & 11 & -1.72 & 19.37 & 0.5179 & RRab \\
OGLE050516.31-684514.9   & 339 & 30 & 35 & -1.79 & 19.41 & 0.5815 & RRab \\
OGLE050516.31-684514.9   & 332 & 33 & 20 & - & 19.41 & 0.5815 & RRab \\
OGLE050519.07-684522.5   & 337 & 14 & 64 & - & 18.41 & 0.4556 & RRab \\
OGLE050542.14-684529.5  & 314 & 36 & 17 & - & 19.42 & 0.2679 & RRe \\
OGLE05054491-6844330 & 319 & 31 & 29 & -1.84 & 19.3 & 0.0000 & RRab \\
OGLE05055814-6846186 & 304 & 13 & 30 & - & 19.08 & 0.4897 & RRab \\
OGLE051731.19-692717.7 & 376 & 13 & 24 & -1.62 & 18.93 & 0.5908 & RRab \\
OGLE051738.47-692628.8 & 309 & 33 & 19 & -1.79 & 19.15 & 0.6341 & RRab \\
OGLE051751.38-692707.5 & 118 & 15 & 35 & -1.61 & 18.89 & 0.5229 & RRab \\
OGLE051751.97-692741.4 & 248 & 3 & 32 & -1.50 & 19.50 & 0.5888 & RRab \\
OGLE051752.63-692856.4 & 252 & 23 & 37 & -1.67 & 19.00 & 0.6663 & RRab \\
OGLE051755.55-692716.3 & 294 & 27 & 59 & -1.56 & 18.33 & 0.5738 & RRab \\
OGLE051755.91-693005.5 & 279 & 17 & 20 & -1.51 & 19.22 & 0.4873 & RRab \\
OGLE051756.52-692934.7 & 413 & 28 & 37 & -1.55 & 19.29 & 0.3532 & RRc \\
OGLE051756.59-692803.2 & 225 & 22 & 44 & -1.53 & 18.74 & 0.4076 & RRc \\
OGLE051808.73-692813.4 & 256 & 23 & 27 & -1.55 & 18.98 & 0.4000 & RRc \\
OGLE051813.74-692840.6 & 228 & 23 & 41 & -1.54 & 19.16 & 0.6497 & RRab \\
OGLE051815.14-692924.1 & 233 & 16 & 21 & -1.55 & 19.28 & 0.5306 & RRab \\
OGLE051825.24-692848.6 & 236 & 24 & 22 & - & 19.44 & 0.4735 & RRab \\
OGLE051825.32-692651.3 & 105 & 28 & 31 & -1.59 & 19.29 & 0.5996 & RRab \\
OGLE051831.26-693054.5 & 155 & 35 & 25 & - & 19.15 & 0.2743 & RRe \\
OGLE051834.08-693049.0 & 188 & 26 & 20 & - & 19.18 & 0.2743 & RRc \\
OGLE051835.21-692913.5 & 296 & 29 & 21 & - & 19.26 & 0.2707 & RRe \\
OGLE051842.29-692642.8 & 277 & 32 & 13 & - & 20.16 & 0.2901 & RRc \\
OGLE052002.59-691431.1 & 244 & 20 & 30 & -1.52 & 18.96 & 0.3365 & RRc \\
OGLE052003.29-691316.1 & 298 & 16 & 32 & -1.52 & 18.89 & 0.5725 & RRab \\
OGLE052005.99-691313.3 & 180 & 20 & 21 & -1.52 & 19.06 & 0.7525 & RRab \\
OGLE052011.82-691424.0 & 335 & 18 & 25 & -1.50 & 19.4 & 0.2639 & RRe \\
OGLE052013.42-691153.0 & 195 & 23 & 33 & -1.53 & 18.97 & 0.6175 & RRab \\
OGLE052018.99-691527.0 & 321 & 37 & 22 & -1.62 & 19.51 & 0.5739 & RRab \\
OGLE052021.61-691235.9 & 211 & 22 & 20 & -1.57 & 19.28 & 0.3314 & RRc \\
OGLE052022.42-691202.3 & 271 & 28 & 11 & -1.52 & 19.33 & 0.7720 & RRab \\
OGLE052026.29-691815.0 & 224 & 14 & 41 & -1.50 & 19.12 & 0.3702 & RRc \\
OGLE052027.12-691607.1 & 181 & 34 & 32 & -1.53 & 19.18 & 0.5844 & RRab \\
OGLE052030.65-691345.2 & 199 & 22 & 21 & -1.50 & 19.10 & 0.3494 & RRc \\
OGLE052032.64-691701.6 & 240 & 28 & 24 & -1.60 & 19.33 & 0.3330 & RRc \\
OGLE052033.36-691334.2 & 284 & 25 & 19 & -1.54 & 19.23 & 0.5805 & RRab \\
OGLE052039.40-691709.0 & 283 & 26 & 18 & -1.52 & 19.05 & 0.2761 & RRe \\
OGLE052043.39-691742.7 & 331 & 23 & 20 & -1.57 & 19.34 & 0.4794 & RRab \\
OGLE052044.57-691726.3 & 258 & 33 & 17 & -1.54 & 19.32 & 0.3497 & RRc \\
OGLE052811.62-693423.8  & 321 & 38 & 29 &-   & 19.50 & 0.5385 & RRab \\
OGLE052822.65-693529.9 & 304 & 26 & 26 & -1.64 & 19.42 & 0.5210 & RRab \\
OGLE052822.98-693310.6  & 319 & 27 & 28 & -1.62 & 19.40 & 0.5800 & RRab \\
OGLE052827.10-693347.7  & 198 & 38 & 14 & -1.47 & 20.06 & 0.4195 & RRc \\
OGLE052829.20-693201.1  & 149 & 20 & 39 & -1.35 & 19.30 & 0.6023 & RRab \\
OGLE052829.28-693442.0  & 252 & 23 & 33 & -1.43 & 19.80 & 0.5755 & RRab \\
OGLE052834.72-693326.4  & 149 & 16 & 36 & -1.58 & 19.37 & 0.6055 & RRab \\
OGLE052854.41-693403.8  & 184 & 28 & 34 & -1.32 & 19.40 & 0.6214 & RRab \\
OGLE053156.19-700954.7 & 234 & 15 & 65 & - & 18.72 & 0.5739 & RRab \\
OGLE053159.01-700959.3 & 228 & 30 & 32 & - & 19.31 & 0.5069 & RRab \\
OGLE053411.38-702413.4 & 220 & 33 & 29 & -1.54 & 19.4 & 0.3400 & RRc \\
OGLE053416.61-702149.6 & 287 & 25 & 41 & -1.51 & 19.26 & 0.3315 & RRc \\
OGLE053418.91-702309.6 & 308 & 38 & 19 & -1.60 & 19.59 & 0.5188 & RRab \\
OGLE053427.44-702249.8 & 335 & 22 & 21 & -1.62 & 19.48 & 0.4764 & RRab \\
OGLE053437.49-702806.9 & 180 & 32 & 22 & - & 19.25 & 0.4757 & RRab \\
OGLE053443.15-702543.8 & 271 & 22 & 23 & -1.50 & 19.39 & 0.2948 & RRe \\
OGLE053459.49-702232.9 & 263 & 18 & 25 & -1.61 & 19.35 & 0.5473 & RRab \\
OGLE053500.75-702533.0 & 277 & 21 & 24 & -1.53 & 19.48 & 0.5014 & RRab \\
OGLE053502.87-702829.6 & 332 & 29 & 32 & -1.50 & 19.36 & 0.3286 & RRc \\
OGLE053504.52-702444.9 & 277 & 24 & 16 & -1.68 & 19.35 & 0.5958 & RRab \\
OGLE053509.25-702145.7 & 248 & 21 & 23 & -1.55 & 19.28 & 0.3587 & RRc \\
OGLE053510.95-702729.9 & 224 & 30 & 7 & - & 21.00 & 0.6274 & RRab \\
OGLE053518.04-702755.2 & 301 & 33 & 15 & - & 19.35 & 0.5134 & RRab \\
OGLE053713.07-695956.8  & 270 & 21 & 12 & -1.10 & 19.98 & 0.6161 & RRab \\
OGLE053714.30-695846.0  & 278 & 28 & 17 & -1.38 & 20.00 & 0.6502 & RRab \\
OGLE053715.52-695812.2  & 214 & 28 & 15 & -1.41 & 20.08 & 0.5353 & RRab \\
OGLE053715.59-700105.8  & 251 & 34 & 5 & -  & 21.00 & 0.5642 & RRab \\
OGLE053717.79-695923.9  & 178 & 23 & 32 & -1.60 & 19.57 & 0.2869 & RRc \\
OGLE053720.10-700202.5  & 291 & 33 & 31 & -1.39 & 19.51 & 0.7042 & RRab \\
OGLE054527.55-710615.3 & 329 & 25 & 11 & - & 19.89 & 0.7787 & RRab \\
OGLE054600.10-711155.3 & 302 & 31 & 28 & -1.71 & 19.31 & 0.5711 & RRab \\
\hline
\end{longtable}
\clearpage

\end{document}